\documentclass[useAMS,usenatbib]{mn2e}
\usepackage{graphics,epsfig,amsmath,amssymb,amstext,txfonts,array,color}
\usepackage{graphicx}
\usepackage[T1]{fontenc}

\usepackage{color}

\usepackage{braket, hyperref}
\usepackage[utf8x]{inputenc}
\usepackage[table]{xcolor}
\usepackage{tabu}
\usepackage{ulem}

\newcommand\degree{$^{\circ}$}

\newcommand\simlt{\lower.5ex\hbox{$\; \buildrel < \over \sim \;$}}
\newcommand\simgt{\lower.5ex\hbox{$\; \buildrel > \over \sim \;$}}

\title[Cosmic-Ray Anisotropy from Large Scale Structure and the effect of magnetic horizons]
{Cosmic-Ray Anisotropy from  Large Scale Structure and the effect of magnetic horizons }

\author[N.~Globus et al.]
  {N.~Globus,$^1$\thanks{E-mail: noemie.globus@mail.huji.ac.il}
  T.~Piran,$^1$  
  Y.~Hoffman,$^1$
  E.~Carlesi$^2$, 
  D.~Pomar\`ede$^3$  \\
  $^1$Racah Institute of Physics, The Hebrew University of Jerusalem, 91904 Jerusalem, Israel\\
  $^2$Leibniz Institut f\"{u}r Astrophysik, An der Sternwarte 16, 14482 Potsdam, Germany\\
  $^3$Institut de Recherche sur les Lois Fondamentales de l'Univers, CEA Saclay, F-91191 Gif-sur-Yvette, France}
  \date{Released \today}

\pagerange{\pageref{firstpage}--\pageref{lastpage}} \pubyear{2002}

\def\LaTeX{L\kern-.36em\raise.3ex\hbox{a}\kern-.15em
    T\kern-.1667em\lower.7ex\hbox{E}\kern-.125emX}

\begin{document}
\label{firstpage}
\maketitle

\begin{abstract}
{Motivated by the $\sim7\%$ dipole anisotropy in the distribution of ultra-high energy cosmic-rays (UHECRs) above 8 EeV, we explore the anisotropy induced by the large scale structure,  using constrained simulations of the local Universe and taking into account the effect of magnetic fields.  
The value of the intergalactic magnetic field (IGMF) is critical as it determines the UHECR cosmic horizon. { We calculate the UHECR  sky maps  for different values of the IGMF variance and  show the effect of the UHECR horizon on the observed anisotropy.}
The footprint of the local ($\lesssim350$ Mpc) Universe on the UHECR background, a small angular scale enhancement in the Northern Hemisphere, is seen.} At 11.5 EeV (the median value of the energy bin at which the dipole has been reported), the LSS-induced dipole amplitude is $A_1\sim10\%$, for IGMF in the range [0.3-3] nG for protons, helium and nitrogen, compatible with the rms value derived from the cosmic power spectrum. {    However at these energies the UHECRs are also influenced by the Galactic Magnetic Field (GMF) and we discuss its effect on the LSS-induced anisotropy. 
}
\end{abstract} 

\section{introduction}\label{intro}

The origin of the ultra-high energy cosmic-rays (UHECRs) is still unknown. To identify a source we need to know the arrival direction of the UHECRs. However, 
UHECRs are deflected on their way to the Earth by the intervening Galactic and { intergalactic} magnetic fields (GMF and IGMF, respectively).  
The only observed statistically significant deviation from isotropy  is a large scale dipole anisotropy \citep[][{\color{black}hereafter PAO17}]{2017Sci...357.1266P}, { of the order of a few percent},
reported at $\sim$5$\sigma$ significance level  for UHECR energies  $E>$8 EeV. 

{ We can expect that extragalactic UHECR sources follow, up to a biasing factor, } the large scale structure (LSS) of the Universe.
Both the energy and composition of the cosmic-rays change during the extragalactic propagation because of their interaction with the cosmological photons backgrounds (GZK effect).  
Moreover, contrarily to the photons, neutrinos and gravitational waves,  UHECRs are deflected by the IGMF, and enter a diffusion regime after a time of a few $D/c^2$ ($D$ is the diffusion coefficient).  

The  observed   UHECR dipole anisotropy is set by the size of the UHECR observable Universe. The "cosmic-ray horizon", the largest distance that the UHECR can propagate {\it at a given energy}, depends  on their diffusion coefficient in the IGMF and on their mean free path in the photons backgrounds {\color{black}\citep{2000PhRvL..84.3527F,2004NuPhS.136..169P, 2010arXiv1005.3311P, 2014PhRvD..89l3001H,2015PhRvD..92f3014H}}. Different  nuclei don't experience the same energy losses, and therefore, even if they have the same rigidity $\sim E/Z$ (i.e. they behave the same way in the IGMF), they have different horizons.  This situation is unique to UHECRs: different energy and nuclei species  probe different distances. Therefore, it has been suggested that, at a given energy and composition,  the  anisotropy in the UHECR background probes  the  {source}  distribution within the cosmic-ray horizon \citep{Waxman+97}.
We investigate  this possibility, assuming that the distribution of UHECR sources follow the LSS. We calculate the UHECR dipole anisotropy induced by the matter distribution, taking  into account the diffusive propagation of the UHECRs in the IGMF.
We derive the amplitude and direction of the 
UHECR dipole, for different IGMF values and different compositions. {   We then estimate the effect of the GMF on the LSS-induced UHECR anisotropy, for proton and nitrogen at 11.5 EeV.}

\begin{figure*}
\centering
\includegraphics[scale=0.5]{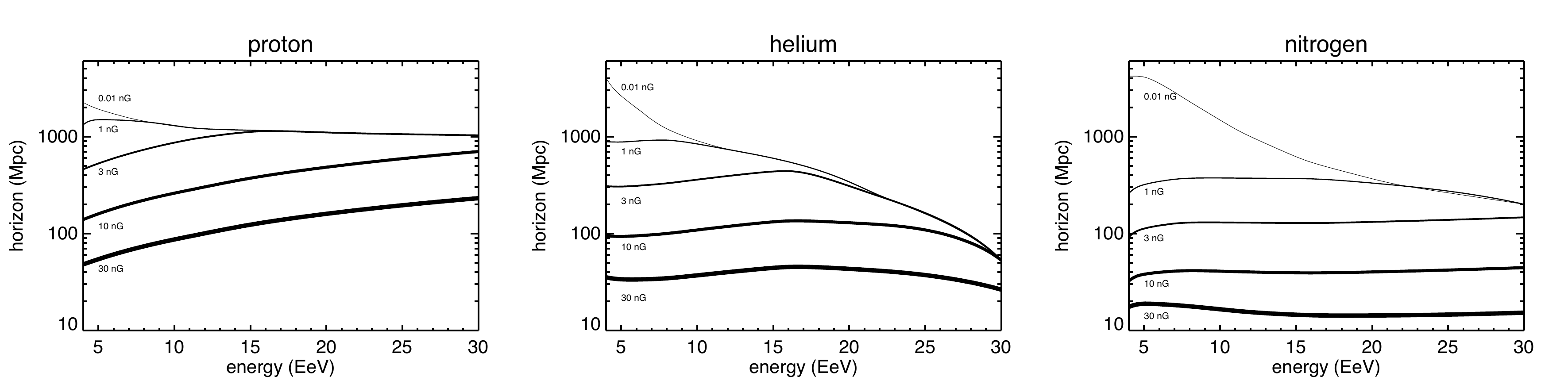}
\caption{UHECR horizons as a function of the energy for different values of the IGMF.}	
\label{horizons}
\end{figure*}

In a previous study \citep[][hereafter GP17]{2017ApJ...850L..25G}, we derived the expected UHECR extragalactic dipole from the observed LSS density power spectrum. We found a maximum value for the rms dipole amplitude of $\sim8b$\% for IGMF strength $\gtrsim1$ nG, for helium and nitrogen at energies greater than 8 EeV. Here $b$ is the bias factor. It is larger than unity if the UHECR sources are more clustered than the dark matter. {\color{black} We showed that the energy dependence of the dipolar amplitude increases as a function of the energy. This is consistent with the findings  by \citet{2014PhRvD..89l3001H,2018ApJ...854L...3W}.} 
The novelty of our approach here is  the  reconstructed density field of the local Universe \citep{2018NatAs.tmp...91H} based on the {\it CosmicFlow-2} catalog of peculiar velocities to calculate sky maps of the UHECR  anisotropy induced by the LSS for different UHECR horizons. {\color{black}  Previous studies used either the 
2 Micron All-Sky Redshift Survey (2MRS)  galaxy catalog for the source distribution \citep{2014PhRvD..89l3001H,2015PhRvD..92f3014H} or the  local large-scale mass structure model of \citet{2005JCAP...01..009D} \citep{2018ApJ...854L...3W} in which the data on the source distribution  extent only to $\sim 110$ Mpc.  The simulations that we use allow to extrapolate the LSS to regions that are poorly observed because
of Galactic foregrounds, and also to probe larger distances, up to $\sim 350$ Mpc. }
 
A 3D view of the density field used in our calculations {\color{black}from \citet{2018NatAs.tmp...91H}} can be explored at this link     \href{https://skfb.ly/6AFxT}{[https://skfb.ly/6AFxT]}. All the major overdensities of the Local Universe are shown as isosurfaces of different colors.

{ To derive the LSS-induced anisotropy, we consider  
 different cosmic-ray horizons that are determined by the diffusion of UHECRs of different compositions in different IGMFs .} 
 
{  In this work we assume a homogeneous and purely turbulent IGMF, {\color{black} and we vary its strength to probe different magnetic horizons}. In reality the IGMF variance is expected to be correlated with the different structures, clusters, filaments, voids
 \citep[e.g.][]{2008PhRvD..77l3003K}. This may change the propagation of the UHECRs in the field with stronger deflections within regions of stronger magnetic field and weaker ones. The overall effect might be mimicked by varying the coherence length or by more detailed simulations. However, as we are mostly interested in the dipole this would probably won't have a significant effect, as already shown by \citet{2018MNRAS.475.2519H} who tested different magnetogenesis models and obtain similar UHECR anisotropy for the different models. Note, however, that these authors didn't consider the GMF which,  as we show later, has a significant effect on the dipole and hence must be taken it into account. Note also that these authors  {\color{black} chose a finite number of source positions up to 140 Mpc, and these positions (as well as the magnetic field structure)  then were periodically repeated, which lead to a rather isotropic distribution} beyond 140 Mpc, reducing the dipole for UHECRs whose horizon is larger than this distance.}
 
 { The plan of the paper is as follow.} We present the {  observations} in section  \ref{observations}. We discuss the effects of the intergalactic magnetic fields on the UHECR horizons in section  \ref{magnetic}. { In section  \ref{Method} we discuss the reconstruction of the density field and the calculations of the UHECR propagation in the IGMF. }  We present the results in section  \ref{results}. {   We show the effect of the GMF in section \ref{Galactic}} and discuss    our results in 
 \ref{discussion}.

\section{The observed UHECR dipole}\label{observations}
The Pierre Auger Observatory (hereafter Auger) has measured a large-angular scale dipolar anisotropy in the distribution of the arrival directions of the ultra-high energy cosmic-rays (UHECR) at energies { above } 8 
 EeV {(with a mean energy of 11.5 EeV)}. 
The dipole amplitude is $A_1 =(6.5^{+1.3}_{-0.9})$\% and its direction  
 ($l$, $b$) = (233\degree,~-13\degree) $\pm10$\degree$\,$ in Galactic coordinates {\color{black}(PAO17)}. 

{ At lower energies, 4-8 EeV (with a mean 5 EeV) Auger  reported a dipole amplitude $A_1 =(2.5^{+1.0}_{-0.7})$\% and a direction ($l$, $b$) = (286\degree,~-32\degree) $\pm10$\degree ($\sim50$\degree$\,$  away from the { $>8$ EeV}  dipole). However this lower energy dipole is not statistically significant. }
 
 The composition seems  to be different in these two energy bins. The interpretation of the measurements of the composition-dependent\footnote{It is also energy-dependent and it should be kept in mind that this dependence is not fully understood yet.} observable $X_{\rm max}$ (the atmospheric depth of the air shower maximum) is as follow \citep{2017JCAP...04..038A}:
 
 Above 8 EeV, the spread in the $X_{\rm max}$ distribution indicates that the composition seems to be dominated by a single component (i.e. one specie dominates at a given energy), likely helium or nitrogen. Specifically,  the EPOS-LHC model \citep{EPOS1,EPOS2}  gives nitrogen-like elements up to $\sim60$ EeV. At the highest energies, i.e. above 60 EeV, the model suggests a composition heavier than nitrogen \citep{2017JCAP...04..038A}, but due to small number statistics the uncertainties are large. 
 
 In the [4-8] EeV energy { range} (where the sky is compatible with isotropy) the mean value of the $X_{\rm max}$ distribution indicates that the composition is dominated by lighter elements.  
The spread in the $X_{\rm max}$ distribution indicates that there is   a mixture of many components at these energies. Specifically, protons { (a significant fraction, up to 50\%)} but also {     heavier nuclei}  seems to be present in this lower energy range.

\begin{figure*}
\includegraphics[scale=0.53]{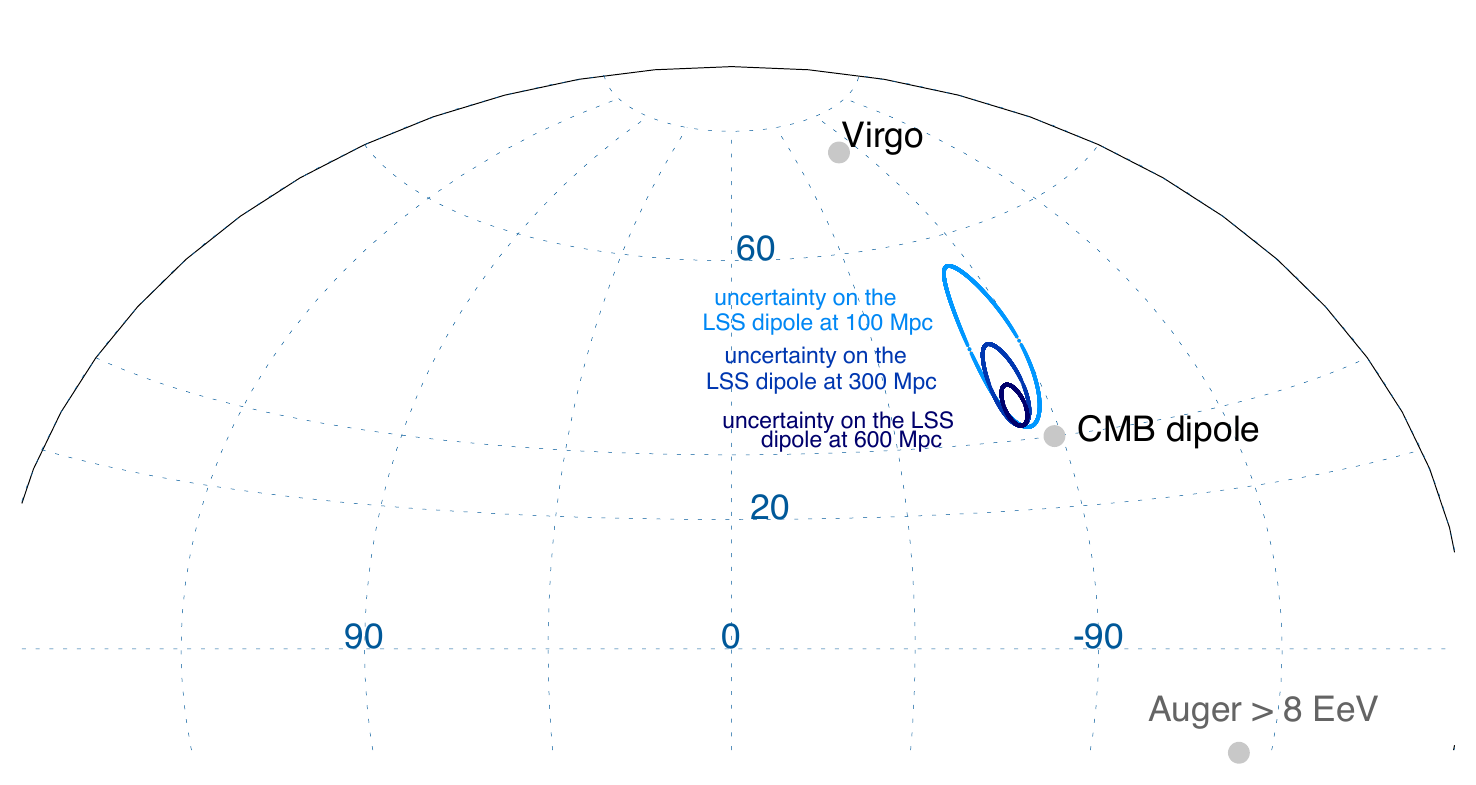}
\includegraphics[scale=1.27]{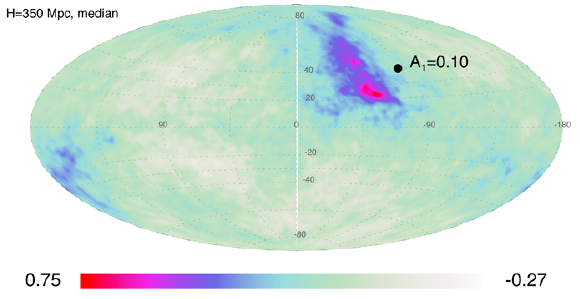}
\includegraphics[scale=0.45]{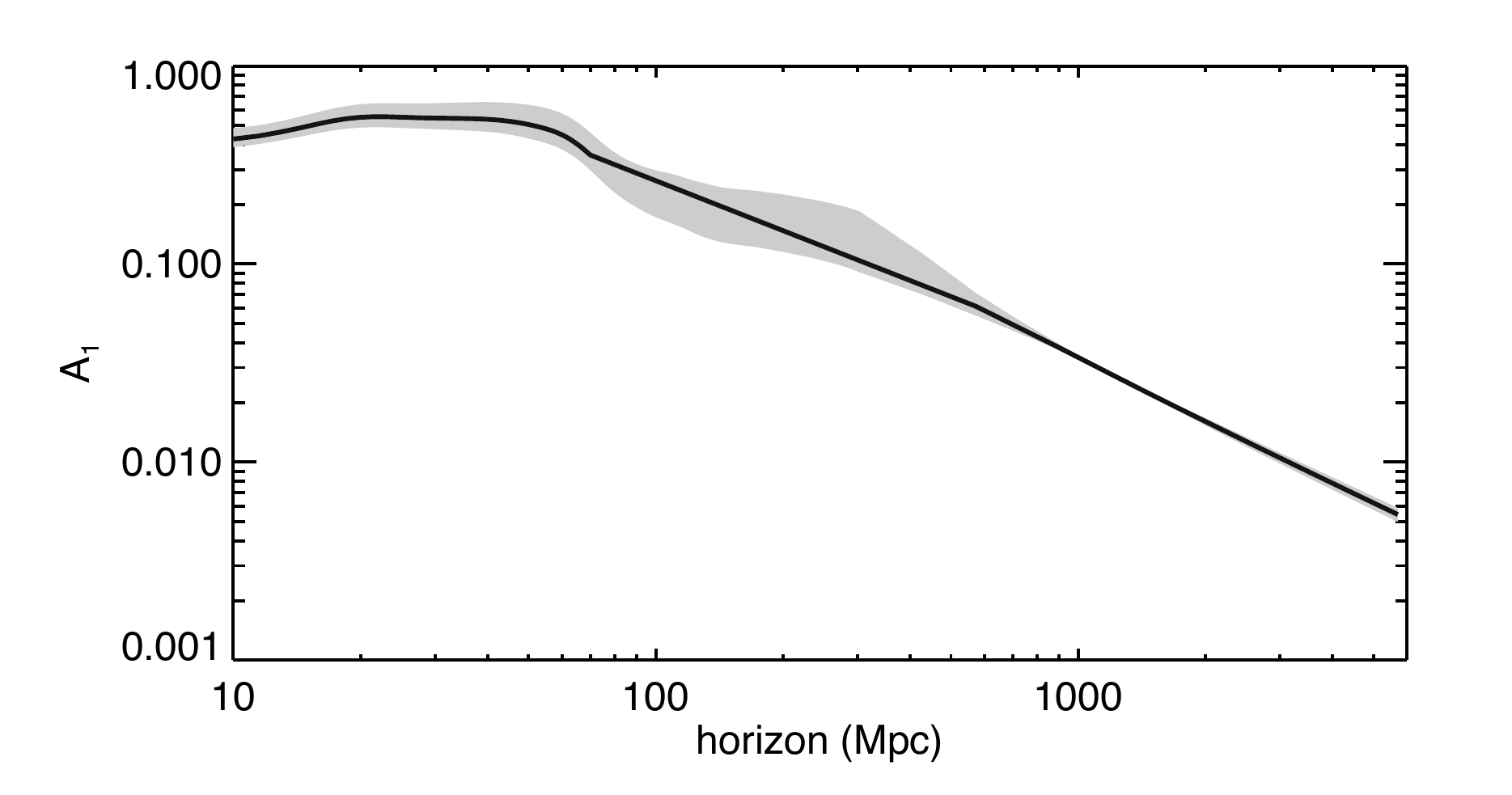}
\caption{{\color{black} Upper left panel:} Directions of the dipole induced by the matter distribution derived in \citet{2018NatAs.tmp...91H}, at different depths. The  QL density field  is shown { at the link }\href{https://skfb.ly/6AFxT}{[https://skfb.ly/6AFxT]}. 
  {\color{black} Upper right panel:} A sky map, in Galactic coordinates,  showing the anisotropy resulting from the integrated density field (column density) up to a maximal distance of 350 Mpc. The amplitude $A_1$ and direction of the dipole induced by this matter distribution is indicated by a black dot.  {\color{black} Bottom} panel: The amplitude of the dipole induced by the matter distribution at different depths.  {   The uncertainty corresponds to the 16 and 84 percentiles of the distribution of the reconstructed density field.}}
\label{QL2}
\end{figure*}

\section{Cosmic-ray horizons, from the ballistic to the diffusive regime}\label{magnetic}
The observable UHECR Universe is limited to the cosmic-ray {\it horizon}, i.e. the distance that a cosmic-ray, at a given energy, can propagate  from its source.  {With no IGMF the UHECR horizon is {\color{black} in good approximation the mean total attenuation length, $d_{\rm GZK}=c(-d\ln E/dt)^{-1}$, 
 including the  contribution of pair production and photodisintegration processes \citep[e.g.][]{2012APh....39...33A}. We use the  new  giant dipole resonance  cross-sections    discussed  in \citet{2005APh....23..191K} and for  the  higher-energy  processes, the  parameterisation  of  \citet{Rachen96}.  }

} 
For a purely turbulent IGMF, {the UHECRs diffuse over  a diffusion distance $d_{\rm diff} \sim 6 D/c$ \citep[e.g.][]{GAP08}
where $D$ is the diffusion coefficient. }

 For a Kolmogorov turbulence, the diffusion coefficient $D$ is well approximated by a fitting function taking into account both the resonant and non-resonant diffusion regimes \citep{GAP08},
\begin{equation}
D\approx0.03\left(\frac{\lambda_{\rm Mpc}^2 E_{\rm EeV}}{ZB_{\rm nG}}\right)^{{1}/{3}}+0.5\left(\frac{E_{\rm EeV}}{ZB_{\rm nG}\lambda_{\rm Mpc}^{0.5}}\right)^{2} {\rm Mpc^2 \,Myr^{-1}}
\label{dcoef}
\end{equation}
where $Z$ is the charge of the cosmic-ray,   $E_{\rm EeV}$ is its energy measured in EeV, 
$B_{\rm nG}$ the IGMF strength in nG and $\lambda_{\rm Mpc}$ its coherence length in Mpc.

{If $d_{\rm diff}$ is smaller than $d_{\rm GZK}$ then the horizon from which the UHECRs can reach Earth becomes: $ \sim{(6D d_{\rm GZK}/c)^{1/2}}$ . }
Combined the UHECR  horizon is given by:
\begin{equation} H(E) =\min(\sqrt{ d_{\rm diff} d_{\rm GZK}} ,d_{\rm GZK}) \ . 
\label{eq:H}\end{equation} 
{While the diffusion in the magnetic field depends just on the rigidity of the nuclei, its GZK distance depends on its type and  energy. Hence different UHECRs will have different horizons even if they have the same rigidity or the same energy. }   
 
Fig.  \ref{horizons}  depicts the horizon distance for { protons, helium, nitrogen as a function of the energy for different IGMF strengths (0.01, 1, 3, 10 and 30 nG) and a maximum turbulence scale of 1 Mpc. 
}
The horizons are  constant (equal to $d_{\rm GZK} $) for small magnetic fields and they decrease linearly with $B$ for higher values. 

\section{Methods}\label{Method}

If the UHECR source distribution follows the LSS, its density is proportional  to the matter density field $\rho(r,\hat{e})$. Anisotropies in the UHECR background are related to the density contrast field $\delta(r,\hat{e})\equiv(\rho(r,\hat{e})-\bar{\rho})/\bar{\rho}$, where  $\hat{e}$ is an arbitrary direction in the sky, and $r$ is the radial distance. We begin by estimating this field.

\subsection{ The Density Field} 

{
In the standard model of cosmology, the matter density and  velocity fields are closely related by the continuity equations. Hence observations of the peculiar velocities of galaxies allow for an unbiased mapping of the underlying mass distribution. The reconstruction of the density field from galaxy peculiar velocities commenced with the POTENT algorithm \citep{1990ApJ...364..349D}. The more powerful and versatile Bayesian approach of the Wiener filter (WF) and constrained realizations (CRs) of Gaussian fields was applied almost a decade later  to reconstruct the LSS from peculiar  velocities \citep{1999ApJ...520..413Z}. 
The WF/CRs methodology provides a better control over the resolution of the recovered LSS and allows for extrapolation of the density and velocity fields to regions outside the data zone (e.g. the Galactic zone of avoidance).  Yet, like the POTENT, the WF/CRs algorithm is strictly valid only in the linear regime, where deviations from the homogeneity and isotropy are small. The WF/ CRs has been applied to the currently state-of-the-art CosmicFlows database of peculiar velocities \citep[][and references therein.]{2014Natur.513...71T, 2017NatAs...1E..36H}. 
The WF/CRs methodology has been applied to set up constrained initial conditions for cosmological simulations, resulting with so-called constrained simulations 
\citep[for a review see][]{2014NewAR..58....1Y, 2014MNRAS.437.3586S}.

The linear WF/CRs algorithm has been recently extended to the quasi-linear (QL) regime by means of constrained simulations \citep{2018NatAs.tmp...91H}. 
 An ensemble of 20 cosmological simulations constrained by the CosmicFlows-2 data \citep{2013AJ....146...86T} was constructed and used to sample to sample the posterior distribution of the present density and velocity fields given the $\Lambda$CDM cosmology and the CosmicFlows-2 data. Taking the mean and variance over this sample provides an estimator for the present epoch QL structure of the nearby universe. The QL density field is estimated by means of the geometric mean of the ensemble of the constrained simulations and the arithmetic mean over the velocity fields provides a proxy to the QL velocity fields. The effective resolution of the QL fields is roughly 5 Mpc. It should be noted that the resolution of the individual constrained simulations is higher but the smoothing induced by the averaging process renders the resolution to  $\sim$5 Mpc. The main effect of the the averaging process is to wash out the internal virial structure of groups and clusters. 

The QL density field is calculated on a Clouds-in-Cells grid of $512^3$ size within a periodic box of $\sim350$ Mpc depth.  
{\color{black}  It is based on an ensemble of 20 CRs. A  view of this median density field is shown at \href{https://skfb.ly/6AFxT}{[https://skfb.ly/6AFxT]}.   
The density field beyond the box boundaries is obtained in the linear regime using a series of linear constrained realisations  \citep[based on the linear WF/CRs algorithm,][]{1991ApJ...380L...5H,1999ApJ...520..413Z} within a $\sim1830$ Mpc depth. The use of the linear realisations is justified as the contributions to the dipole from beyond the box of $\sim350$ Mpc are dominated by large (linear) scales, deep enough in the linear regime.
The linear and QL realisations sample the posterior probability distribution of the density field given the CosmicFlows-2 data and the assumed standard model of cosmology. Hence mean and scatter of the density field are readily evaluated.
At larger distances we simply assume that the Universe is homogeneous, i.e. $\delta(r,\hat{e})=0$ for $r>1830$ Mpc.    For most cases of interest this is larger than the GZK distance.}

As UHECRs with different energy and composition have different horizons we show in Fig. \ref{QL2} the amplitude and direction of the dipole induced by this  density field at different depths  with no IGMF.  The uncertainty in the density field   leads to an  uncertainty in the dipole amplitude, which is shown  in Fig. \ref{QL2} by the shaded area.
{\color{black} This uncertainty corresponds to the scatter of our 20 CRs of the density field. }
 The uncertainty in the direction  of the dipole due to the uncertainty in the distribution of matter is also shown at different depths (100, 300 and 600 Mpc). At large distances, the direction of the  dipole due to the matter distribution converges to the direction of the CMB dipole. {As we see later, once the horizon distance of a given UHECR is known (see Fig. \ref{horizons}) the overall properties of the corresponding LSS induced dipole can be read from this figure. }

\begin{figure*}
     \rotatebox{0}{\includegraphics[scale=0.5]{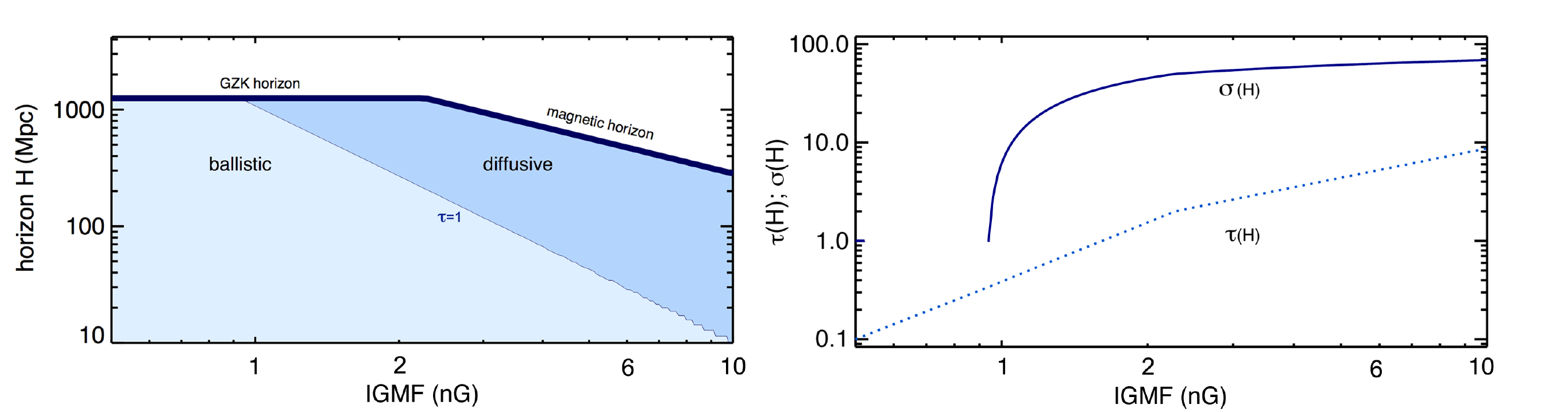}}\\
     $\,$\\
   \includegraphics[scale=0.85]{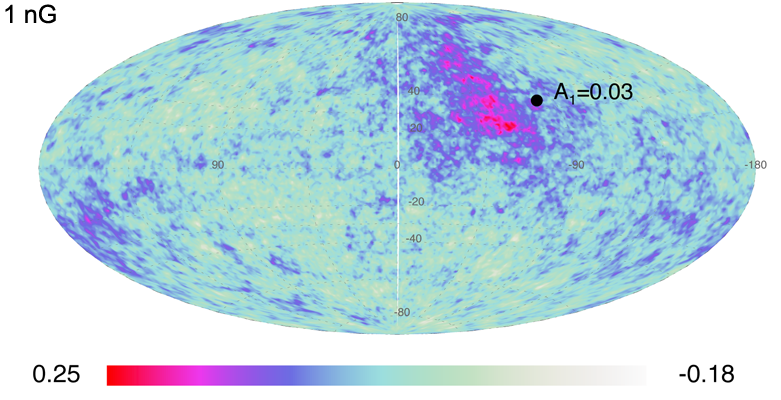}
\includegraphics[scale=0.85]{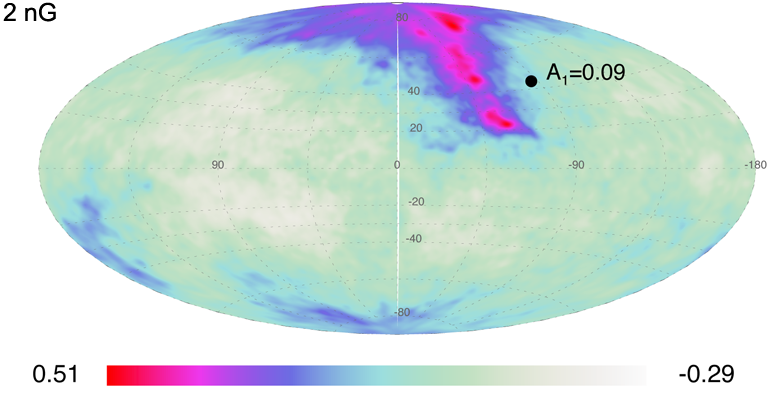}
\includegraphics[scale=0.85]{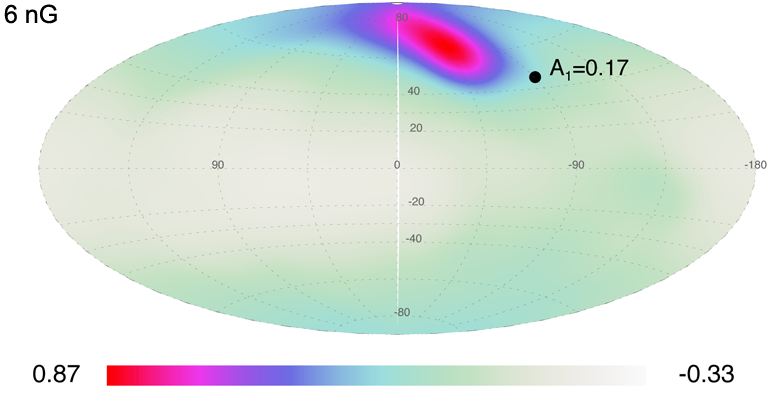}
   \caption{Top: left panel, the magnetic horizon $H$ (i.e. size of the UHECR observable Universe) and  the radius at which  ${\tau=1}$  for 11.5 EeV protons, as a function of the IGMF. Right panel, the optical depth $\tau$ and the angular spread $\sigma$ of a single source at the horizon  as a function of the IGMF. Lower panel: LSS-induced  anisotropy for different IGMF. Three different characteristic behaviour are shown and corresponds to IGMF : 1, 2 and 6 nG. In the left sky map, all the observable Universe is in the ballistic regime, as can be seen on the upper plot. In the middle sky map, the local Universe ($\lesssim350$ Mpc) is still in the  ballistic regime while the rest of the Universe is in the diffusive regime. Right sky map: only very local sources ($\lesssim30$ Mpc) are in the ballistic regime. For a LSS-induced anisotropy this corresponds to a concentration in the direction of the Virgo cluster. The amplitude $A_1$ and direction of the dipole component are indicated by black dots. An animation { showing the evolution of the anisotropy for IGMF variances} from 0.01 nG to 10 nG is available in the supplementary materials.}
\label{p_11EeV}
\end{figure*}

\subsection{Calculation of the UHECR anisotropy}

{ The intensity profile of a cosmic-ray source on the sky depends on
the scattering properties of the particles in the IGMF. These properties depend in turn on the  optical
depth\footnote{ This optical depth is not the optical depth for a single scattering. Instead it corresponds to a large angle scattering $\braket{\delta\theta^2}\sim 1$ that arises from multiple small scatterings.}
\begin{equation}
\tau={rc}/{D}
\label{tau}
\end{equation} 
where $r$ is the distance from the source and on the typical single scattering angle, characterized by the rms value 
\begin{equation}
\braket{\delta\theta^2}\sim(\lambda_c/\beta r_L)^{\kappa}{\braket{\delta B^2}/B^2}\,,
\label{deltatheta}
\end{equation}
where $\kappa=2$ for $\lambda_c\leq r_L$ and 
$\kappa=-2/3$ for $\lambda_c> r_L$ for a Kolmogorov turbulence \citep{2008PhRvD..77l3003K}. We assume strong turbulence ($\braket{\delta B^2}/B^2 =1$).

Once $\tau$ and $\braket{\cos(\delta\theta)}$ are known, the image of the  source is calculated as follow:
using $\tau$ and  $\braket{\cos(\delta \theta)}$,
we estimate the rms angular width of the source $\sigma(\tau,\braket{\cos(\delta \theta)}) = \sigma(r,D)$  \citep{Narasimhan}.
We than approximate the angular distribution of the cosmic-rays from this source as a Gaussian with this width. 
This Gaussian, $G$,  characterizes for a given shell at a distance $r$ and a diffusion coefficient $D$, 
the angular distribution on Earth from sources on this shell. Integrating over all distances up to the cosmic-ray horizon we obtain
\begin{equation}
I_i(\hat{e},D_i(E))=\int_0^{H_i(E)} \int_{\Omega'} \delta(r,\hat{e}') G[{\cos}^{-1}(\hat{e}\cdot\hat{e'})/\sigma(r,D_i(E),\delta\theta_i) ] dr d\Omega'
\end{equation}
where $\delta(r,\hat{e}')$ is the source intensity (we assume it is proportional to the density contrast), $D_i(E)$ is the diffusion coefficient, $\delta\theta_i$ is the rms scattering angle and $H_i(E)$ the cosmic-ray horizon, where the scalar $i$ denotes the particle species. 

At a given energy $E$, the cosmic ray horizon, the diffusion coefficient and the single scattering angle depends on the nature of the particles (section 2) and on the IGMF parameters ($B,\lambda_c$).
We derive sky maps $I(\hat{e},D(E))$, for different composition ($Z=1,2,7,14$) and IGMF values.
The sky maps are shown in Galactic coordinates.
We calculate the dipole moment (amplitude and direction) of each sky map.

}

\section{Results}\label{results}

\subsection{Sky maps}

{ Sky maps of the LSS-induced anisotropy for protons at 11.5 EeV, are shown in Fig.\ref{p_11EeV}. The UHECR horizon, as well as the distance at which the Universe become diffusive to cosmic-rays ($\tau\sim1$) is indicated in the upper panel. {   We also indicate the angular size $\sigma(r,D(E))$ of a source located at the horizon and at a distance $r=350$ Mpc.}} 

{ First in practically all cases there is an enhancement in the direction ($l=310$\degree$\pm10$\degree, $b=40$\degree$\pm25$\degree) }. This corresponds (see Fig. \ref{QL2}) to a direction between the CMB dipole (the distant universe)  and Virgo (the very local universe).  { It is interesting to note that Cen A  is located at the edge of this region and that clustering has been reported at energies  in this direction \citep{Aab15}.}
The dipole is in this basic direction as well, but its  amplitude varies among the different cases. When the UHECR horizon is large ($>350$ Mpc), the dipole direction converges to the direction of the CMB dipole.

Beyond this basic structure one can see { three} different characteristic behaviours in this figure:

1) In cases that magnetic diffusion is unimportant and $d_{\rm GZK}$ is large, the effective horizon is of order Gpc.  UHECR propagation is ballistic and the anisotropy map is granular with a significant small scale structure. The common enhancement in the direction of ($l=310$\degree$\pm10$\degree, $b=40$\degree$\pm25$\degree) appears here as well.  The overall magnitude of the dipole is small, of order of a 3\% percent. 

2) The second case correspond to a situation in which the magnetic diffusion is important.    
One first notice that the anisotropy maps are smoother now, because of the enhanced diffusion. 
In this case we can see the effect of the structure of "local" (i.e. few hundred Mpc) Universe.  Corresponding to the LSS dipole amplitude shown in Fig. \ref{QL2} the dipole amplitude is
of order 10\%, compatible with the rms dipole amplitude  (GP17) and with the observed dipole.

3) The third case correspond to a situation in which most sources are in the diffusion regime. The dipole amplitude is large, of order of 20\%.

We calculated sky maps for protons, nitrogen and silicon at 50 EeV in 1 nG IGMF. Protons at 50 EeV have a sky map similar to the left one in the figure \ref{p_11EeV} and nitrogen and silicon to the middle one.

{   The transitions between the three cases take place at different values of the IGMF for different energy and  nuclei specie. The diffusion coefficient, eq.\ref{dcoef}, as well as the single scattering angle, eq.\ref{deltatheta}, scales with the particle rigidity. The UHECR horizon, equation \ref{eq:H}, depends on the type of nuclei. For a given specie and IGMF values, we can determine the angular spread of a source at the horizon $\sigma_H$. The first sky map corresponds to $\sigma_H\sim1$\degree, the second to $\sigma_H\sim45$\degree, the third one to  $\sigma_H\sim90$\degree.}

\subsection{The LSS-induced dipole anisotropy} \label{XGal}

 There are two effects that controls the dipole amplitude: first, the horizon size (a smaller horizon implies a larger dipole) and second, the diffusion in the IGMF { (a smaller diffusion coefficient implies a smaller  dipole amplitude)}.\\
 1) When the IGMF is negligible,   the amplitude of the dipole anisotropy is {\it constant}, set by the GZK distance. \\
 2) When the IGMF starts to become significant, 
  the amplitude of the dipole {\it increases} because the dominant contribution to the dipole is given by sources located at $r\lesssim350$ Mpc that are still in the ballistic regime. Therefore the effect of decreasing the horizon size is dominant over the diffusion effect.\\
3) For IGMF large enough,  the local Universe ($r\lesssim350$ Mpc) enters the diffusion regime and therefore the amplitude of the dipole {\it decreases} because the dominant effect is the diffusion.

Fig. \ref{compo11} shows the dipole amplitude for different composition at 11.5 EeV. This corresponds to the median value of the energy at which the dipole has been reported.
At 11.5 EeV the composition seems to be dominated by nitrogen (see section \ref{observations}). We find an amplitude of $\sim10\%$  for nitrogen for an IGMF  $\sim$0.3 nG.

Fig. \ref{compo5} shows the dipole amplitude at 5 EeV. 
The lower energy bin seems to be dominated by light elements (see section \ref{observations}). For an IGMF of 0.3 nG we find a  dipole amplitude for protons smaller than $2\%$, compatible with the observed value.

\begin{figure}
\centering
\includegraphics[scale=0.3]{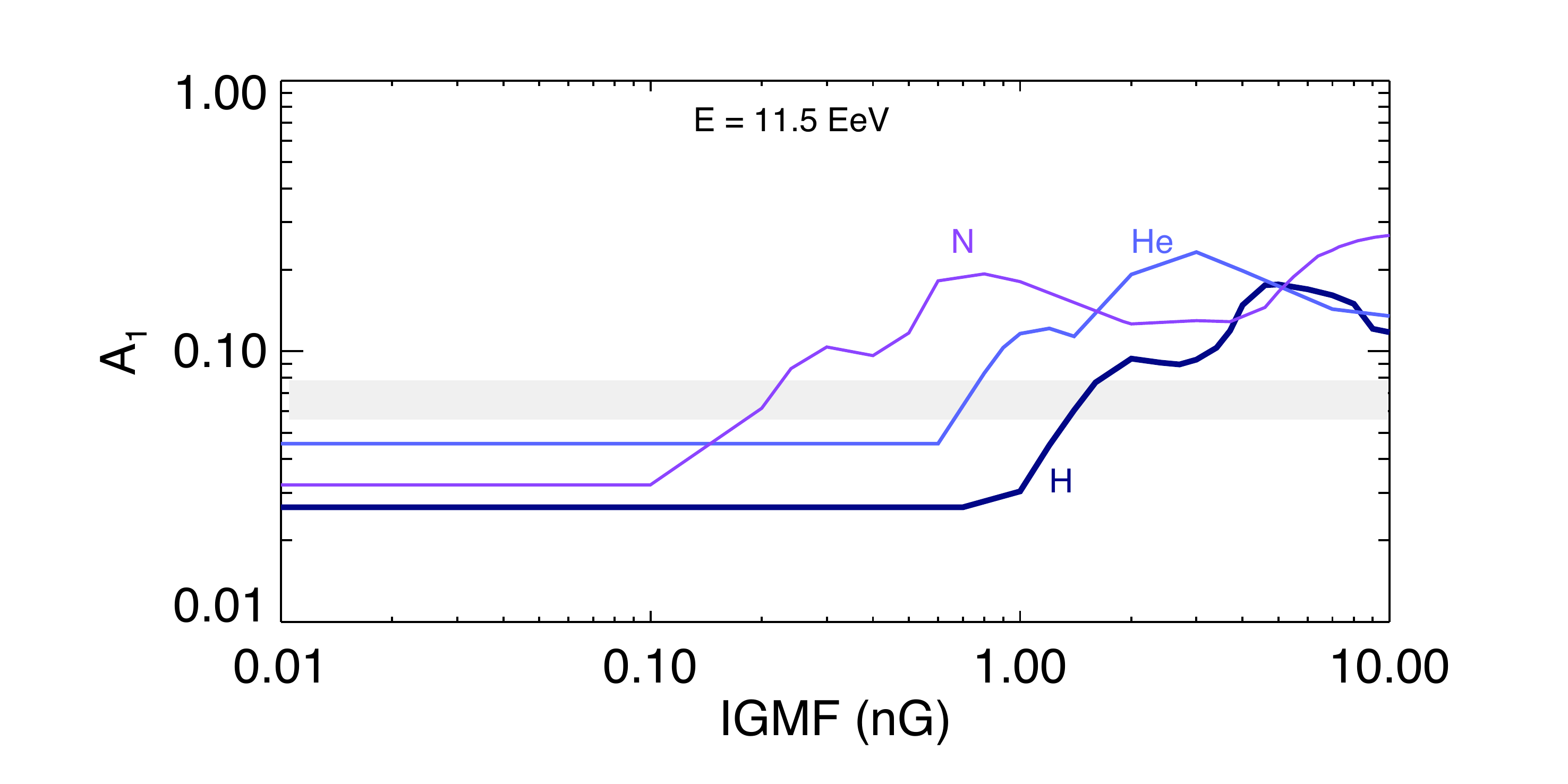}
\caption{LSS-induced dipole at 11.5 EeV for different composition. { While we use the mean of the 20 realizations here, the uncertainty on the LSS-induced dipole can be read on Fig. \ref{QL2}. } The observed amplitude with its uncertainty is figured by the shaded area.}	
\label{compo11}
\end{figure}
\begin{figure}
\centering
\includegraphics[scale=0.3]{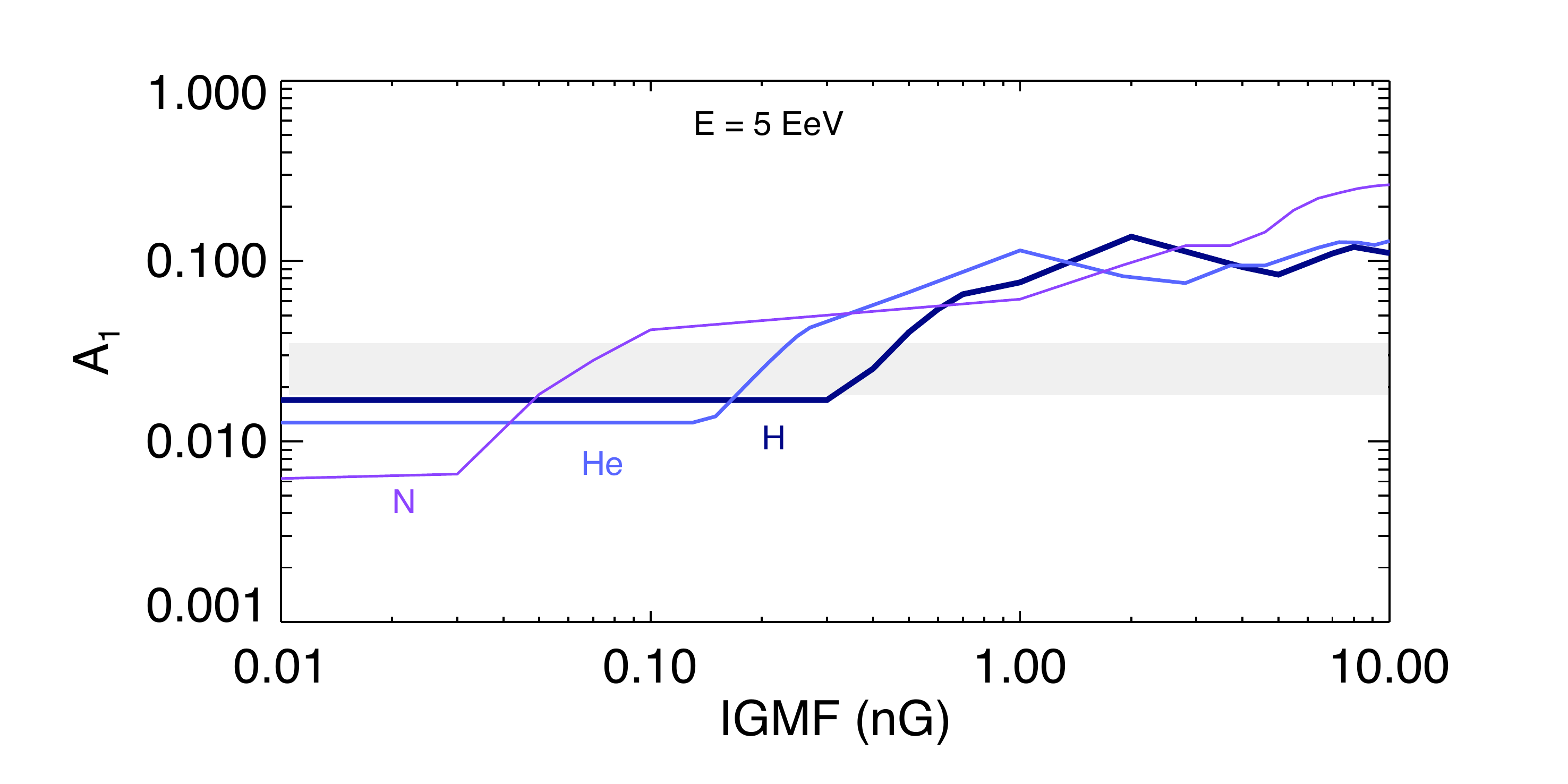}
\caption{LSS-induced dipole at 5 EeV for different composition. The observed amplitude with its uncertainty is figured by the shaded area.}	
\label{compo5}
\end{figure}

\begin{figure*}
\centering
\includegraphics[scale=0.85]{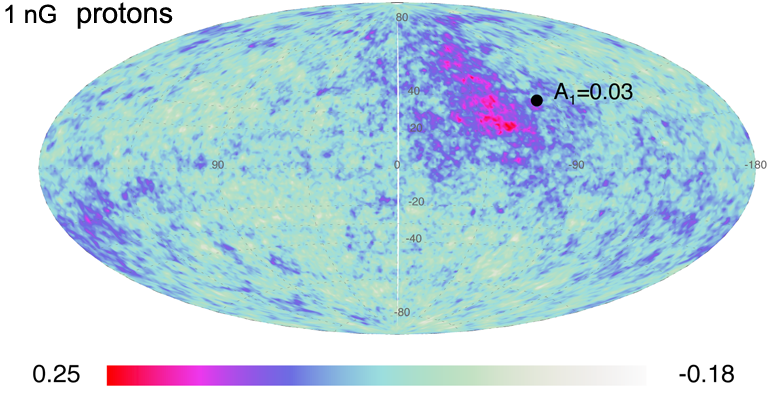}  \includegraphics[scale=0.83]{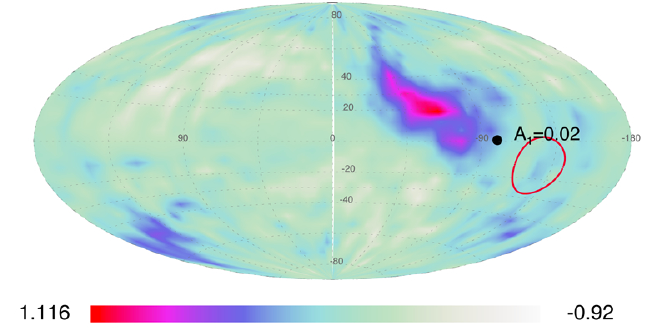}\\
\includegraphics[scale=0.85]{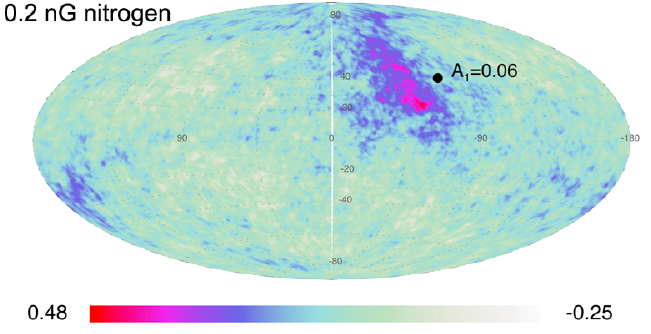}\includegraphics[scale=0.83]{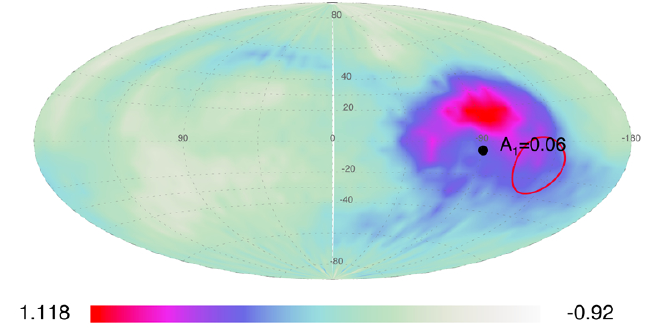}\\
 \includegraphics[scale=0.83]{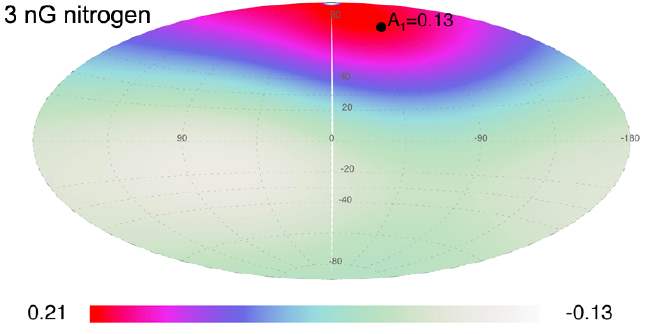} \includegraphics[scale=0.83]{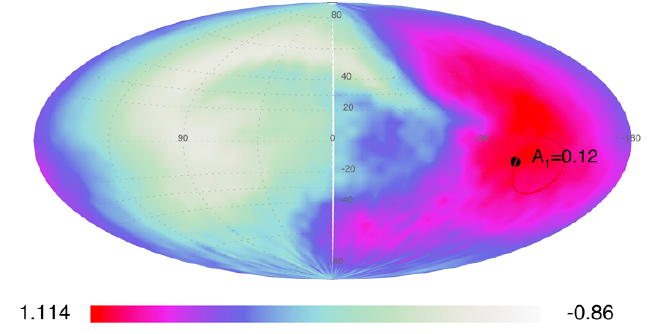}\\
   \caption{Sky maps, in Galactic coordinates, of the LSS-induced UHECR anisotropy taking into account the effect of the Galactic magnetic field of \citet{JF12}. Left, from top to bottom: the LSS-induced UHECR anisotropy for protons at 11.5 Eev in 1 nG IGMF, nitrogen in 0.2 and 3 nG IGMF respectively. Right: the anisotropy after reconstruction by the GMF of \citet{JF12}. The amplitude $A_1$ and direction of the dipole  are marked by the black dot. { The observed Auger dipole direction is figured by the red circle.}
    }
\label{GMF}
\end{figure*}
\section{The effect of the Galactic magnetic field } \label{Galactic}

 {\color{black} Deflections in the GMF were discussed already in PAO17 \citep[see also][]{2016IAUFM..29B.723F}. However PAO17 considered only the regular component of the \citet{JF12} model.
Moreover PAO17 estimated only the deflection from a single direction, {\it i.e.} the dipole in the 2MRS galaxy distribution. Here we reconstruct the sky maps expected from the LSS up to $\sim 350$ Mpc, taking into account the effect of magnetic horizons. The extragalactic dipole direction   depend on the magnetic horizon and does not necessarily coincide with the 2MRS dipole.}
 
 {\color{black} We used the complete GMF model by \citet{JF12,JF12b}  inferred from observations of Faraday rotation, synchrotron emission and polarized dust emission.  
 The coherence length is 20 pc for the isotropic turbulent component and 100 pc for the striated turbulent component. The amplitude of the striated component has been scaled down by a factor 0.3 and the amplitude of the isotropic component by a factor 0.6, that has been shown to be in better agreement with the data \citep{2016JCAP...05..056B}.
 
Alternative models \citep{2011ApJ...738..192P} only fitted to Faraday rotation would give similar results for the dipole amplitude \citep{2018MNRAS.476..715D} but since we are interested here also in the dipole direction, we preferred to use the more advanced model. The overall size of the deflections are comparable in the two models but the direction of the deflections can be  different \citep[see Fig.15 of][]{2018PrPNP..98...85M}. The uncertainties on the anisotropy due to our lack of knowledge of the GMF will be the aim of another study.}

The GMF has a lensing effect that leads to a distortion of the fluctuations of the extragalactic UHECR background. This effect depends on the position of the source in the sky and on the particle's rigidity. 
At rigidities $\gtrsim 10$ EV,  UHECRs are  deflected significantly by the GMF \citep{2016IAUFM..29B.723F}. 

{ To reconstruct the sky maps after GMF propagation, we use the antiparticle tracing method  \citep{Thielheim68}. We back-propagate $2\,10^5$ anti-protons and anti-nitrogen in the GMF magnetic field of \citet{JF12,JF12b}, at a the energy 11.5 EeV. The initial directions ($l_{in},b_{in}$) of the velocity vectors are uniformly distributed. 
 For each ($l_{in},b_{in}$) there is a corresponding direction outside the Galaxy,   ($l_{out},b_{out}$), where the trajectory of the particle becomes ballistic.
Each trajectory is assigned an extragalactic intensity $I(l_{out},b_{out})$ based on the previous calculations. 
We use  $6$\degree pixels for the sky maps on Earth. 
For each pixel on Earth,  we assign the average extragalactic intensity arriving to that pixel. 
The resolution of 6\degree on Earth is sufficient to indicate how the extragalactic anisotropy is deformed by the GMF.
 }

As an illustration of the effect of the GMF, we show in Fig.\ref{GMF}  examples of  sky maps, using the GMF model of \citet{JF12}, of a LSS-induced anisotropy for protons and nitrogen at 11.5 EeV.  
At smaller rigidities, the image is a combined effect of the IGMF and GMF, as can be seen for nitrogen for two different values of the IGMF, 0.2 and 3 nG.

{ The GMF smooths the sky map but also changes the direction of the dipole. The dipole moment of the sky map is moved from the Northern to the Southern hemisphere, and it is not far from the direction observed by Auger.  The effect of changing the  direction of the anisotropy is due to the large scale regular component of the GMF, while the smoothing is due to the turbulent component of the GMF. }

\section{Discussion}\label{discussion}

{ We calculated the LSS-induced UHECR anisotropy, assuming that the source density is proportional to the matter density $\rho$. The novelty is to use constrained simulations from \citet{2018NatAs.tmp...91H} which provide an estimate of the local cosmic { density field up to $\sim350$ Mpc}. We developed an original approach to be able to calculate the sky maps for different IGMFs  and discussed the effect of magnetic horizons on the UHECR anisotropy.  }

With the density field of the local Universe, we recover a dipole amplitude of the same order as the rms value (GP17),  $A_1\sim0.1$, for IGMF in the range [0.3-3] nG for  11.5 EeV protons, helium and nitrogen.  We recall that this energy is the median value of the energy bin at which the dipole has been reported. \citet{2018NatAs.tmp...91H} calculated the bias relation between the matter and the density field, $\rho_g=(\rho/\bar{\rho})^{b}$,  
{ with $b = 1.74  \pm   0.13$ for the luminosity density density field derived from the compilation of the 2M++  redshift survey of  galaxies \citep{2015MNRAS.450..317C}.   
We opted here to make the simplest assumption that the intensity of the UHECR sources follows the mass distribution. Relaxing this assumption and allowing for a linear bias factor implies that the results quoted here about the anisotropy need to be multiplied by this bias factor. }
{\color{black} A significant bias between the matter density and the sources would increase the dipole amplitude.}

The anisotropy induced by the LSS presents small-scale structures.  If protons were dominating at   11.5 EeV, a proton anisotropy would not be significantly altered by the GMF \citep[e.g.][]{2016IAUFM..29B.723F}.  It is interesting to note that the LSS-induced anisotropy  presents an enhancement which is not far from the Cen A direction ($l=310$\degree, $b=20$\degree), and that a clustering was already reported in that direction by Auger  \citep{Aab15}.  
If nitrogen is dominating at 11.5 EeV, then the rigidity is smaller.

 For an IGMF of a few nG, we obtain a large scale angular anisotropy (well represented by a dipole) in the Virgo  direction.  Our preliminary considerations of the GMF suggest that it deflects the LSS-induced dipole towards the direction observed by Auger.

While we have presented here calculations only for a single value of the energy, a full parametric analysis is planned in a further paper to better understand  the effect of varying the IGMF and GMF parameters  for a given evolution of the composition with energy. \\

NG and TP were supported by the I-CORE Program of the Planning and Budgeting Committee,  the Israel Science Foundation (grant 1829/12), and an advanced ERC grant TReX.
YH was supported by an ISF grant 1013/12.    
We  thank the Red Espa\~nola de  Supercomputaci\'on  for granting us computing time in the Marenostrum Supercomputer at the BSC-CNS where the simulations used for  this paper have been performed.

\end{document}